\documentclass[twoside,english]{elsarticle}
\usepackage[T1]{fontenc}
\usepackage[latin9]{inputenc}
\usepackage{array}
\usepackage{dvipost}
\usepackage{textcomp}
\usepackage{amsbsy}
\usepackage{amstext}
\usepackage{graphicx}

\makeatletter

\providecommand*{\perispomeni}{\char126}
\AtBeginDocument{\DeclareRobustCommand{\greektext}{%
  \fontencoding{LGR}\selectfont\def\encodingdefault{LGR}%
  \renewcommand{\~}{\perispomeni}%
}}
\DeclareRobustCommand{\textgreek}[1]{\leavevmode{\greektext #1}}
\DeclareFontEncoding{LGR}{}{}

\newcommand{\lyxmathsym}[1]{\ifmmode\begingroup\def\b@ld{bold}
  \text{\ifx\math@version\b@ld\bfseries\fi#1}\endgroup\else#1\fi}

\providecommand{\tabularnewline}{\\}

\@ifundefined{showcaptionsetup}{}{%
 \PassOptionsToPackage{caption=false}{subfig}}
\usepackage{subfig}
\makeatother

\usepackage{babel}

\begin{document}

\title{The Structural and Magnetic ordering in $La{}_{0.5-x}Nd_{x}Ca_{0.5}MnO_{3}$
(0.1 \ensuremath{\le} x \ensuremath{\le} 0.5) Manganites}

\author[SSPD]{Indu Dhiman}

\author[SSPD]{A. Das}

\ead{adas@barc.gov.in}

\author[TPD]{P. K. Mishra}

\author[ugc]{N. P. Lalla}

\author[SSPD]{and A. Kumar}

\address[SSPD]{Solid State Physics Division, Bhabha Atomic Research Centre, Mumbai
- 400085, India}

\address[TPD]{Technical Physics Division, Bhabha Atomic Research Centre, Mumbai
- 400085, India}

\address[ugc]{UGC-DAE Consortium for Scientific Research, University Campus, Khandwa
Road, Indore 452017, India}
\begin{abstract}
The crystal and magnetic structure of polycrystalline
$La{}_{0.5-x}Nd{}_{x}Ca_{0.5}MnO{}_{3}$ (0.0 \ensuremath{\le} x
\ensuremath{\le} 0.5) samples have been investigated using
magnetization, resistivity, transmission electron microscope, and
neutron diffraction techniques. The samples are isostructural and
possess orthorhombic structure in \textit{Pnma} space group. On
lowering of temperature, the samples exhibit CE - type
antiferromagnetic structure coexisting with a weak ferromagnetic
ordering. The charge and orbitally ordered antiferromagnetic phase
is weakened by the growth of ferromagnetic phase. The evolution of
structural distortions and magnetic structure at low temperature
as a function of Nd doping exhibit a strong correlation with A -
site disorder ($\sigma{}^{2}$).\end{abstract}
\maketitle

\section{Introduction}

Half doped manganites $R{}_{0.5}A_{0.5}MnO_{3}$ (R: a trivalent
earth ion, A: a divalent earth ion) have been attracting
considerable interest due to strong coupling between the charge,
spin, and orbitals. The charge and spin ordering in these
compounds is highly sensitive to several perturbations such as
disorder effects at the rare earth (A \textendash{} site) and
transition metal (B \textendash{} site) site, hydrostatic
pressure, magnetic and electric field, and size of the particles
\cite{C N R Rao,Dagotto,J. B. Goodenough,Y. Tokura}. In
particular, substituting at A \textendash{} site modifies the
average A \textendash{} site ionic radius $<r{}_{A}>$ and
introduces disorder ($\sigma{}^{2}$) which is found to have
significant influence on magnetic and transport properties.
Disorder is expressed as $\sigma{}^{2}=\sum
x_{i}r_{i}^{2}-<r_{A}>^{2}$, where $x{}_{i}$ denotes the
fractional occupancy of the A-site ion and $r{}_{i}$ is the
corresponding ionic radius and $<r{}_{A}>$ is the average A-site
ionic radius \cite{L. M. Rodriguez-Martinez}.

Half doped manganites with $<r{}_{A}>$$\sim1.20-1.263$Å and
disorder $\sigma{}^{2}$$\sim10{}^{-3}-10{}^{-2}$ exhibit stable
ferromagnetic phase coexisting with antiferromagnetic phase.
Modifying the ionic radii with substitutions at the rare earth
site generally leads to a suppression of the ferromagnetic phase.
This is evidenced in the following studies. The influence of
doping in systems such as
$\left(Nd{}_{1-y}Sm_{y}\right){}_{0.5}Sr_{0.5}MnO_{3}$ causes
reduction in both $T{}_{C}$ and $T{}_{CO}$ \cite{H. Kuwahara}. In
Y doped $Pr{}_{0.5}Sr_{0.5}MnO_{3}$ compound, the ferromagnetic
phase is suppressed, while the antiferromagnetic transition
temperature $T{}_{N}$ either remains stable or shows a slight
increase \cite{J. Wolfman}, whereas with Nd doping $T{}_{N}$ shows
a reduction and $T{}_{C}$ remains nearly stable \cite{L. S. Ling}.
Magnetic and transport study of
$La{}_{0.5-x}Ln{}_{x}Sr_{0.5}MnO_{3}$ (Ln = Pr, Nd, Gd and Y) have
shown a change over from ferromagnetic metallic state to an
antiferromagnetic insulating phase with increase in x \cite{Md.
Motin Seikh}. Similar studies on the effect of doping at the rare
earth site in compounds with lower ionic radii $<r{}_{A}>$
$\sim1.14-1.20$Å and disorder
$\sigma{}^{2}$$\sim10{}^{-6}-10{}^{-4}$ have been investigated.
Curiale et al. have studied the effect of Pr doping in
$\left(La{}_{y}Pr{}_{1-y}\right)_{0.5}Ca_{0.5}MnO_{3}$ manganites
and observe a rapid decline in ferromagnetic transition
temperature $T{}_{C}$ and increase of the charge ordering
transition temperature ($T{}_{CO}$) \cite{J. Curiale}. In
$Nd{}_{0.5-x}La{}_{x}Ca_{0.5}MnO_{3}$ and
$Pr{}_{0.5-x}La{}_{x}Ca_{0.5}MnO_{3}$ systems the occurrence of a
reentrant ferromagnetic state on cooling and a marked effect of
disorder on $T{}_{C}$ has been observed \cite{P. V. Vanitha,D.
Zhu,S. Yang}. Most of these studies mainly focus on the influence
of A \textendash{} site disorder on magnetic transitions. The
effect of disorder on the crystal and magnetic structure,
including phase separation phenomenon, has not been studied
widely.

The phase separation behaviour observed in half-doped manganites
has also attracted wide theoretical interest \cite{A.
Moreo,Dagotto-1}. The appearance of a charge ordered CE-type
antiferromagnetic ground state in manganites could not be
reproduced using a superexchange model alone. The theoretical
studies show that electron \textendash{} phonon coupling
($\lambda$) due to cooperative or non-cooperative
Jahn\textendash{}Teller phonons in addition to double exchange and
super exchange effects was necessary to reproduce the phase
diagram of half-doped manganites with charge-orbitally ordered
CE-type antiferromagnetic state \cite{J. van den Brink,L. Brey}.
The random substitution of $La^{3+}$ with $Nd^{3+}$ ions may be
identified with the quenched disorder. In the presence of quenched
disorder the hopping of $e_{g}$ electrons and the exchange
interaction $J_{AF}$ between $t_{2g}$ spins is affected due to the
buckling of Mn\textendash{}O\textendash{}Mn bonds. Taking into
account the fluctuations of hopping and exchange couplings, Monte
Carlo studies reveal the coexistence of giant clusters of
ferromagnetic and antiferromagnetic phases \cite{A. Moreo-1}.
Variation of disorder can lead to preferential stabilization of
one phase as compared to the other. In a similar study, which
included the electron \textendash{} phonon coupling
(\textgreek{l}), CE- and A-type antiferromagnetic and
ferromagnetic phases were realized as a function of $\lambda$ and
$J_{AF}$ \cite{K. Pradhan}.

The effect of A - site disorder on the nature of magnetic ordering
in charge ordered manganites has been reported previously.
However, in these studies the change in <$r{}_{A}$> and
$\lyxmathsym{$\sigma$}^{2}$ were significantly large \cite{A.
Das,P. D. Babu,I. Dhiman,I. Dhiman-1,I. Dhiman-2,C.
Autret-Lambert,S. Savitha Pillai}. In the present study we have
investigated the influence of introducing very small disorder
($\sim10{}^{-4}$) in $La{}_{0.5}Ca_{0.5}MnO_{3}$ compound. Towards
this, we have studied compounds in the series
$La{}_{0.5-x}Nd{}_{x}Ca_{0.5}MnO{}_{3}$ (0.1 \ensuremath{\le} x
\ensuremath{\le} 0.5). The end compound
$La{}_{0.5}Ca_{0.5}MnO_{3}$ (<$r{}_{A}$> = 1.198Å) exhibits
$T{}_{C}$ \ensuremath{\approx} 225K, $T{}_{CO}$
\ensuremath{\approx} $T{}_{N}$ \ensuremath{\approx} 170K, having
CE - type antiferromagnetic ground state with magnetic moment
predominately oriented in the ac - plane \cite{P. G. Radaelli},
whereas in $Nd{}_{0.5}Ca_{0.5}MnO_{3}$ (<$r{}_{A}$> = 1.17Å)
ferromagnetic character is totally suppressed and it exhibits
$T{}_{CO}$ \ensuremath{\approx} 240K, $T{}_{N}$
\ensuremath{\approx} 150K, CE - type antiferromagnetic ground
state with moments oriented along a or c axis \cite{F. Millange}.
We find that introducing even a small disorder by susbstituting La
with Nd in these class of compounds results in pronounced
enhancement of micrometer size ferromagnetic clusters coexisting
in an antiferromagnetic matrix.

\section{Experimental Details}

The polycrystalline samples were synthesized by conventional solid
state reaction method. The starting materials $La{}_{2}O{}_{3}$,
$MnO{}_{2}$, $Nd_{2}O{}_{3}$, and $CaCO{}_{3}$ were mixed in
stoichiometric ratio and fired at 1200\textdegree{}C for 48hrs.
Samples were then repeatedly ground and heated at
1400\textdegree{}C for 48hrs. Finally, the samples were pelletized
and sintered at 1400\textdegree{}C for 60hrs. Initial values of
cell parameters for all the samples were obtained from X-ray
powder diffraction at 300K with a Rigaku diffractometer, rotating
anode type using Cu K$\alpha$ radiation. Neutron diffraction
patterns were recorded on a multi PSD based powder diffractometer
($\lambda$ = 1.249Å) at Dhruva reactor, Bhabha Atomic Research
Centre, Mumbai at selected temperatures between 22 and 300K, in
the 5\textdegree{} \ensuremath{\le} 2$\theta$ \ensuremath{\le}
140\textdegree{} angular range. The powdered samples were packed
in a cylindrical Vanadium container and attached to the cold
finger of a closed cycle Helium refrigerator. Rietveld refinement
of the neutron diffraction patterns were carried out using
FULLPROF program \cite{J. Rodriguez-Carvajal}. Neutron
depolarization measurements (\textgreek{l} = 1.205Å) were carried
out on the polarized neutron spectrometer at Dhruva reactor,
Bhabha Atomic Research Centre, Mumbai with $Cu{}_{2}MnAl$ (1 1 1)
as the polarizer and $Co{}_{2}Fe$ (2 0 0) as analyzer. For
transmission electron microscopy (TEM) studies the samples were
prepared using conventional technique of ion milling at 3.3KV at
3\textdegree{} grazing incidence of the two Ar ion-guns. A
$LN{}_{2}$ based double-tilt sample holder (636MA-Gatan) was used
for carrying out low temperature TEM measurements, between 98 and
300K. The magnetization measurements were carried out using a
SQUID magnetometer (Quantum Design, USA). The zero field (ZFC) and
field cooled (FC) measurements were carried out in an external
magnetic field H = 0.5T. Standard four probe technique was used to
measure the dc resistivity between 3 and 300K.

\section{Results and Discussion}

\subsection{Crystal structure and distortions}

The Nd doped samples in the series
$La{}_{0.5-x}Nd{}_{x}Ca_{0.5}MnO{}_{3}$ (0.1 \ensuremath{\le} x
\ensuremath{\le} 0.5) are isostructural and possess orthorhombic
structure \textit{\emph{in}}\textit{ Pnma} space group. These
compounds have perovskite structure with O\textasciiacute{}
orthorhombic insulating phase characterized by b/\textsurd{}2
\ensuremath{\le} a \ensuremath{\le} c. In this structure
cooperative Jahn \textendash{} Teller distortions are superimposed
on the rotation of $MnO{}_{6}$ octahedra \cite{B. B. van Aken,G
Maris}. Volume exhibits nearly linear reduction as a function of
Nd doping and is ascribed to a smaller ionic radius of $Nd{}^{3+}$
ions in comparison to $La{}^{3+}$ ions \cite{R. D. Shannon}. The
structural parameters obtained from Rietveld refinement of neutron
diffraction patterns at 300 and 22K are summarized in table I and
II, respectively. The disorder albeit small, displays a maximum
for intermediate composition in the range of x = 0.2 - 0.3.

\begin{figure}[t]
\vspace{0 cm}
 \resizebox{0.50\textwidth}{!}{
\includegraphics{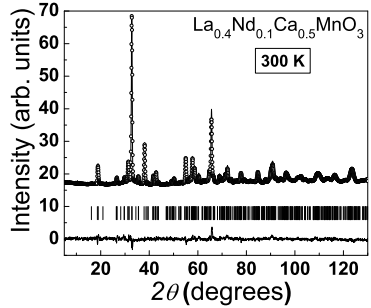}

}\caption{\label{Nd01}A representative neutron diffraction pattern
for $La{}_{0.4}Nd{}_{0.1}Ca_{0.5}MnO{}_{3}$ (x = 0.1) sample at
300K. The continuous line through the data points is the Rietveld
fit to the nuclear phase in \textit{Pnma} space group. The
vertical tick marks below the pattern corresponds to the indexing
of the nuclear peaks in orthorhombic structure and curve at the
bottom shows the difference between observed and calculated
intensity. }

\end{figure}

\begin{figure}[t]
\vspace{0 cm}
 \resizebox{0.50\textwidth}{!}{
\includegraphics{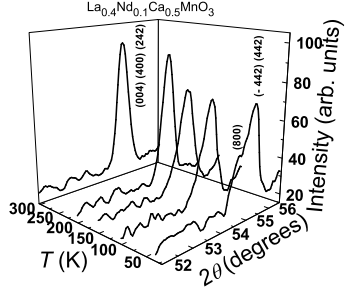}}\caption{\label{Nd01selected}A selected portion (51.5\textdegree{} $\leq$
$2\theta$ $\leq$ 56\textdegree{}) of neutron diffraction patterns
at various temperatures for
$La{}_{0.4}Nd{}_{0.1}Ca_{0.5}MnO{}_{3}$ (x = 0.1) sample is shown.
Splitting of (0 0 4) (4 0 0) (2 4 2) peak in \textit{Pnma} space
group to ($\overline{4}$ 4 2) (4 4 2) and (8 0 0) in $P2{}_{1}/m$
space group is evident at 22K, marking the orthorhombic to
monoclinic structural transition.}

\end{figure}

Figure \ref{Nd01} shows the neutron diffraction pattern for x =
0.1 sample at 300K and Rietveld fit to the data. The values of
lattice parameters obtained are consistent with the ones obtained
from refinement of x \textendash{} ray diffraction patterns at
300K and those reported in literature for the end compounds
\cite{P. G. Radaelli,F. Millange}. On lowering of temperature
below the charge ordering temperature $T{}_{CO}$, a structural
transition to a lower symmetry monoclinic structure in space group
$P2{}_{1}/m$ is expected, as has been reported in similar systems
exhibiting charge ordering behavior \cite{P. G. Radaelli}. We
observe a signature of this in the form of splitting of nuclear
Bragg reflections (2 4 2) (0 0 4) (4 0 0) below $T{}_{CO}$, as
shown in figure \ref{Nd01selected}. The selected portion of
neutron diffraction pattern between 51.5\textdegree{}
\ensuremath{\le} $2\theta$ \ensuremath{\le} 56\textdegree{} show
that the splitting of nuclear Bragg reflections (2 4 2) (0 0 4) (4
0 0) in \textit{Pnma} space group to ($\overline{4}$ 4 2) (4 4 2)
and (8 0 0) in $P2{}_{1}/m$ space group occurs below 150K. This
signifies the transformation of orthorhombic phase in
\textit{Pnma} space group to charge and orbitally ordered
monoclinic phase in $P2{}_{1}/m$ space group. In $P2{}_{1}/m$
space group $Mn{}^{3+}$ and $Mn{}^{4+}$ ions occupy two distinct
sites, in accordance with charge order scenario proposed in the
Goodenough model \cite{J. B. Goodenough-1}. However, the debate
surrounding the charge disproportionation in the charge ordered
state, between Goodenough \cite{J. B. Goodenough-1} and Zener
polaron \cite{A. Daoud-Aladine,L. Wu} model has not been settled
yet in the literature. In the absence of high resolution data, we
obtain similar values for $\chi{}^{2}$ and R factors on fitting
the diffraction pattern in both $P2{}_{1}/m$ and \textit{Pnma}
space groups. The lowering of symmetry to $P2{}_{1}/m$ space group
requires refinement of 29 positional parameters. This reduces the
reliability of refined positional parameters. Therefore, the low
temperature crystal structure is refined in \textit{Pnma} space
group having 7 positional parameters, which yields an average
structure as has been done in the case of x = 0 compound using
neutron diffraction data \cite{P. G. Radaelli}. In samples x = 0.2
and 0.3 with higher Nd doping, splitting attributable to the
orthorhombic to monoclinic structural transition is reduced. This
possibly indicates the modification of charge and orbital ordering
in Nd doped samples. The TEM measurements discussed below display
the presence of incommensurate charge ordering below 150K for x =
0.3 sample as against commensurate charge ordering observed for x
= 0 sample \cite{C. H. Chen}. In the Nd - rich end, the
orthorhombic to monoclinic transition again becomes evident below
200 and 225K for x = 0.4 and 0.5 samples, respectively.

\begin{figure}[t]
\vspace{0 cm}
 \resizebox{0.50\textwidth}{!}{
\includegraphics[scale=2]{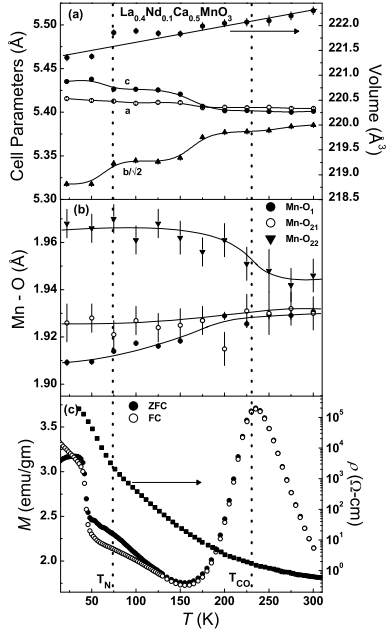}

}\caption{\label{T dependNd01}\textbf{(a)} Temperature dependence
of lattice parameters and the unit-cell volume \textbf{(b)} Bond
lengths as a function of temperature and \textbf{(c)} Variation of
magnetization and resistivity as a function of temperature for
$La{}_{0.4}Nd{}_{0.1}Ca_{0.5}MnO{}_{3}$ (x = 0.1) compound is
shown. The continuous lines are a guide for the eye. }

\end{figure}
\begin{figure}[t]
\vspace{0 cm}
 \resizebox{0.50\textwidth}{!}{
\includegraphics[scale=2]{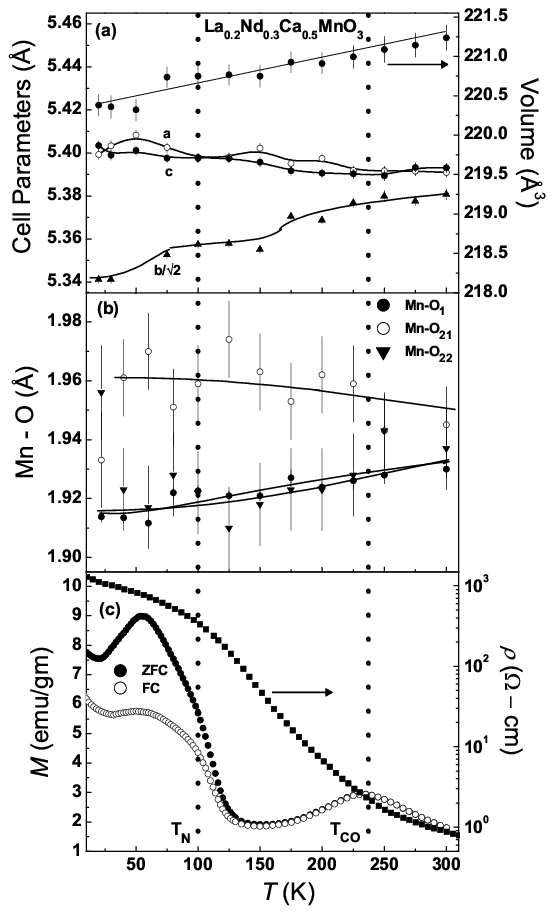}

}\caption{\textbf{\label{T dependNd03}(a)} Temperature dependence
of lattice parameters and the unit-cell volume\textbf{ (b)} Bond
lengths as a function of temperature and \textbf{s} Variation of
magnetization and resistivity as a function of temperature for
$La{}_{0.2}Nd{}_{0.3}Ca_{0.5}MnO{}_{3}$ (x = 0.3) sample is shown.
The continuous lines are a guide for the eye. }

\end{figure}
\begin{figure}[t]
\vspace{0 cm}
 \resizebox{0.50\textwidth}{!}{
\includegraphics[scale=2]{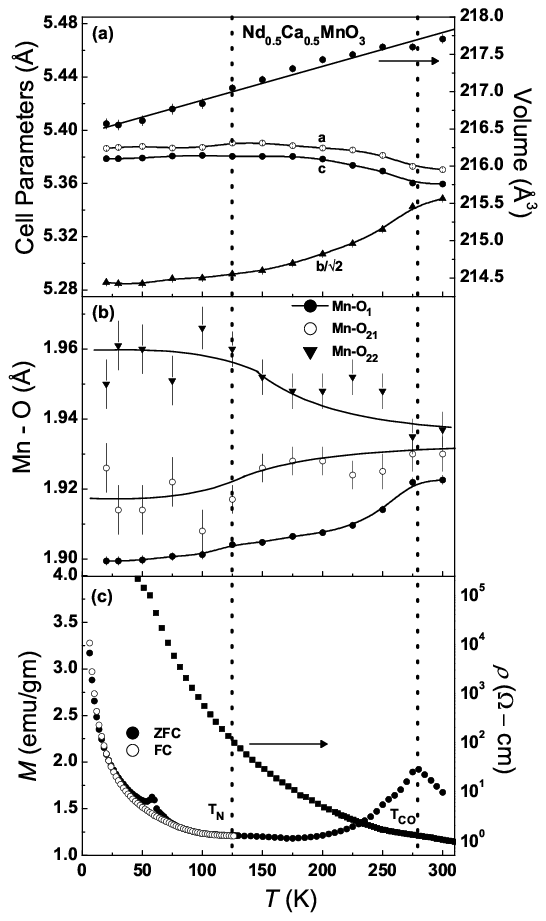}

}\caption{\textbf{\label{T dependNd05}(a)} Temperature dependence
of lattice parameters and the unit-cell volume \textbf{(b)} Bond
lengths as a function of temperature and \textbf{(c)} Variation of
magnetization and resistivity as a function of temperature for
$Nd{}_{0.5}Ca_{0.5}MnO{}_{3}$ (x = 0.5) sample is shown. The
continuous lines are a guide for the eye. }

\end{figure}
The temperature evolution of structural parameters, magnetization
and resistivity for x = 0.1, 0.3, and 0.5 samples are shown in
figures \ref{T dependNd01}, \ref{T dependNd03}, and \ref{T
dependNd05}, respectively. As temperature is reduced below 300K,
the cell parameters, a and c exhibit a slight increase while b
decreases drastically. This anomalous behavior of lattice
parameters has also been observed in $La{}_{0.5}Ca_{0.5}MnO{}_{3}$
(x = 0) and is associated with the ordering of $d{}_{z^{2}}$
orbitals in the a-c plane \cite{P. G. Radaelli}. Similar behavior
of cell parameters as a function of temperature is observed in
samples with x \ensuremath{\le} 0.4. The temperature dependent
variation of lattice parameters in $Nd{}_{0.5}Ca_{0.5}MnO{}_{3}$
(shown in figure \ref{T dependNd05}(a)) exhibits behavior similar
to samples with x \ensuremath{\le} 0.4, although the lattice
parameter a > c at all temperatures below 300K, in agreement with
the previously reported study \cite{F. Millange}. The Mn-O bond
lengths as a function of temperature for samples x = 0.1, 0.3, and
0.5 are shown in figure \ref{T dependNd01}(b), \ref{T
dependNd03}(b), and \ref{T dependNd05}(b), respectively. In
samples with x \ensuremath{\le} 0.2, the difference between three
Mn \textendash{} O bond lengths increases as temperature is
reduced, indicating increase in distortion. However, in x = 0.3
sample, the Mn \textendash{} O bond lengths exhibit no significant
change as a function of temperature and remain very close to each
other, indicating the suppression of JT distortions with the onset
of ferromagnetic ordering. At higher doping with x
\ensuremath{\ge} 0.4, Mn \textendash{} O bond lengths display
behavior similar to samples with x \ensuremath{\le} 0.2. The
temperature dependence of Mn-O bond lengths for sample x = 0.5 is
shown in figure \ref{T dependNd05}(b), and it exhibits behavior
similar to x \ensuremath{\le} 0.2 samples. In contrast, the
Mn-O-Mn bond angles remain nearly constant as a function of
temperature in all the samples. To quantify the Mn-O bond length
distortion, $\triangle{}_{JT}$ =
$\frac{1}{3}\sum\left[\left(d_{n}-d\right)/d_{n}\right]^{2}$
\cite{J. Rodriguez-Carvajal} as a function of Nd doping at 22K is
calculated and are given in table II. The bond length distortion
$\text{\ensuremath{\triangle}}{}_{JT}$ exhibits a minimum at x =
0.3 compound. The minimum in $\triangle{}_{JT}$ coincides with the
minimum evident in charge ordering transition temperature
($\text{T}{}_{CO}$, obtained from the hump in magnetization) and
resistivity values at 50K, for x = 0.3 sample. This indicates a
close correlation between $\text{\ensuremath{\triangle}}{}_{JT}$,
magnetic and transport properties and is in agreement with the
results from a similar study on the effect of Y doping in
$Nd{}_{2/3}Ca_{1/3}MnO{}_{3}$ \cite{E. Fertman}.

\subsection{Magnetic and Transport Studies}

The temperature dependence of magnetization, M(T), under zero
field (ZFC) and field cooled (FC) conditions for x = 0.1, 0.3, and
0.5 samples at magnetic field H = 0.5T, are shown in figures
\ref{T dependNd01}(c), \ref{T dependNd03}(c), and \ref{T
dependNd05}(c), respectively. The Nd doped samples undergo
multiple magnetic transitions as a function of temperature. On
reducing temperature below 300K a peak in magnetization at T =
235K is observed in the case of x = 0.1 sample. This coincides
with the onset of charge ordered insulating transition and $T_{C}$
as observed in x = 0 \cite{P. G. Radaelli-1}. On further reduction
of temperature a shoulder at T = 100K occurs, which we identify
with antiferromagnetic ordering temperature,
$T\mbox{\ensuremath{_{N}}}$. The nature of antiferromagnetic
ordering below 100K is studied by neutron diffraction and is
discussed below. Upon further cooling, M exhibits a sharp increase
indicating another ferromagnetic transition at T = 45K. Although
from neutron diffraction study no evidence for the presence of
long range ferromagnetic ordering is observed in x = 0.1 sample. A
magnetic transition coinciding with this temperature has been
reported in $Nd{}_{1/2}Ca_{1/2}MnO{}_{3}$ sample, attributed to
canted antiferromagnetic state \cite{T. Vogt,P. Murugavel}. The
canting angle in the present compounds may be too small to reflect
changes in the neutron diffraction experiments. However,
ferromagnetic nano clusters coexisting with the majority
antiferromagnetic phase is also another possibility for the
ferromagetic behavior in the antiferromagnetic state \cite{J. C.
Loudon}. In addition, a bifurcation between ZFC and FC curves is
observed at 120K. In all the samples with x \ensuremath{\le} 0.3,
M(T) exhibits an anomalous behavior, the value of M in ZFC mode is
higher than FC mode. Similar behavior has been observed in other
charge ordered manganites, attributed to the development of
ferromagnetic ordering below $T{}_{C}$ that does not get
completely suppressed in the antiferromagnetic state. Therefore,
the system is blocked into a metastable state, a feature of the
first order transition and phase separated state \cite{L.
Ghivelder }. We understand that this phenomena is intrinsic to
systems exhibiting charge ordering behavior. The transition to a
charge ordered state is accompanied by a change in structure and
therefore the resulting distortions are different across the
transition. The difference between the ZFC and FC states
essentially arises from application of magnetic field in
crystallographically different structures. Magnetic field
dependent dielectric studies carried out in charge ordered systems
exhibit significant influence of magnetic field on the dielectric
behavior, suggesting that the distortions are influenced by the
magnetic field \cite{C. R. Serrao}. The peak temperature in M(T)
identified as $\text{T}{}_{CO}$ shows a minimum at x = 0.3
composition, coinciding with the minimum observed in the
$(\triangle_{JT})$ distortion parameter. Furthermore, at very low
temperature below 20K an appreciable increase in M(T) is observed.
Such a behavior has been ascribed to a magnetic contribution due
to $Nd{}^{3+}$ sublattice. This becomes stronger and in the case
of $Nd{}_{1/2}Ca_{1/2}MnO{}_{3}$ this tail extends to a much
higher temperature \cite{F. Millange}. At higher Nd concentration
(x \ensuremath{\ge} 0.4), the difference in M(T) for ZFC and FC
mode is considerably reduced, indicating the absence of
ferromagnetic character. This is in agreement with previously
reported magnetization study on $Nd{}_{1/2}Ca_{1/2}MnO{}_{3}$
compound \cite{F. Millange}.

\begin{figure}[t]
\resizebox{0.50\textwidth}{!}{ \subfloat[x = 0.2 at 10 and
140K]{\includegraphics{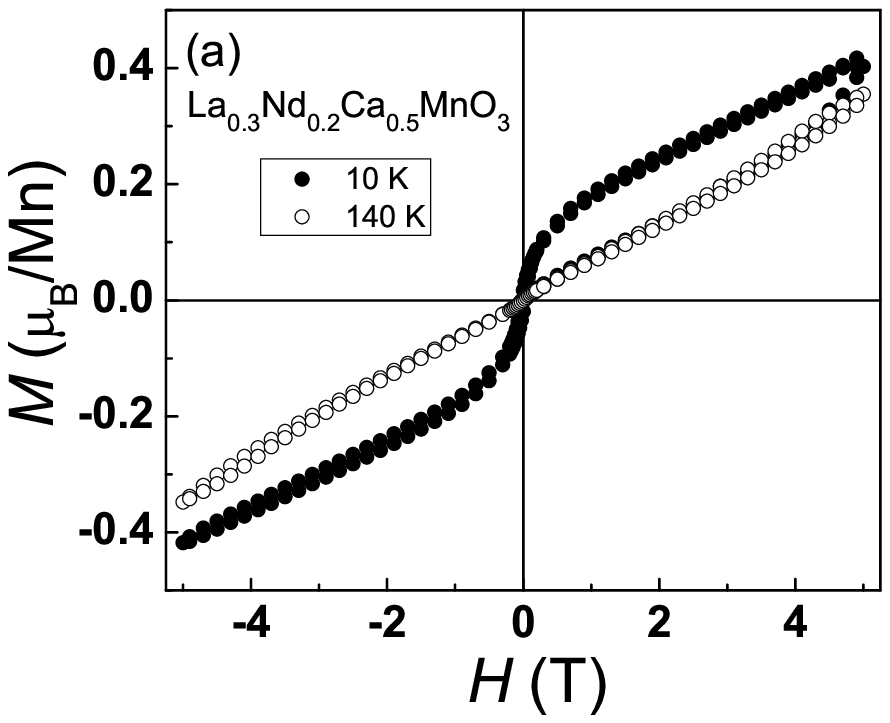}}}\resizebox{0.50\textwidth}{!}{
\subfloat[for x = 0.3 at 5 and 150K]{\includegraphics{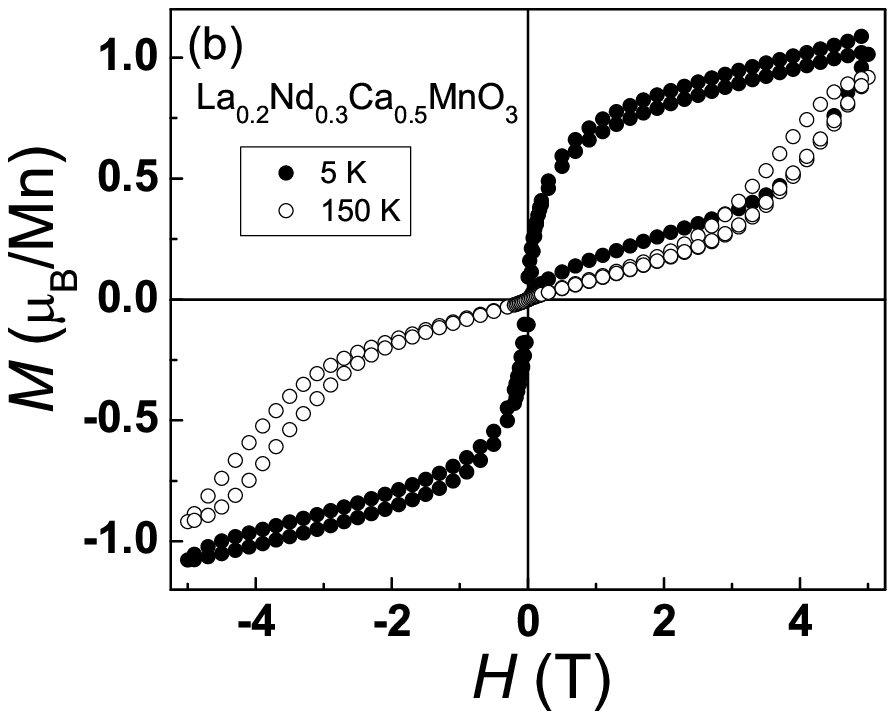}}
}\caption{\label{M(H)}Variation of magnetization (M) with field
(H) for $La{}_{0.5-x}Nd{}_{x}Ca_{0.5}$\-$MnO{}_{3}$ series having
\textbf{(a)} x = 0.2 at 10 and 140K and \textbf{(b)} for x = 0.3
at 5 and 150K.}

\end{figure}
\begin{figure}[t]
\resizebox{0.50\textwidth}{!}{
\includegraphics{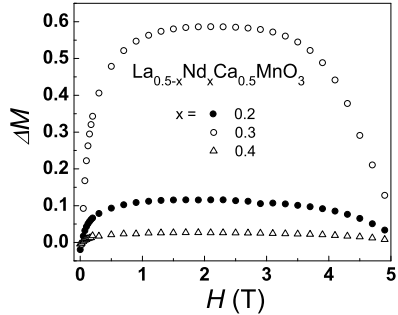}

}\caption{\label{Diff-M(H)}Difference of magnetization on the
envelope and virgin curve as a function of field for
$La{}_{0.5-x}Nd{}_{x}Ca_{0.5}MnO{}_{3}$ series having x = 0.2,
0.3, and 0.4.}

\end{figure}
The variation of magnetization with field, M(H), for samples x =
0.2 and 0.3 are shown in figures \ref{M(H)}(a) and (b),
respectively. At low temperatures (10K for x = 0.2 and 5K for x =
0.3) the M(H) displays a narrow hysteresis loop and a large slope
at high fields in M(H), suggesting the coexistence of
ferromagnetic and antiferromagnetic phases. Additionally,
anomalous behavior in the form of virgin curve lying outside the
envelop curve for both x = 0.2 and 0.3 and a step like behavior
only for x = 0.3 sample is visible. The steps are sharp at low
temperatures and shifts to higher fields and becomes less distinct
with increase in temperature. High resolution neutron diffraction
studies carried out in presence of field have found that the step
in M(H) in charge ordered systems with CE type antiferromagnetic
structure is accompanied by a change in the cell parameters
\cite{C. Yaicle,V. Hardy}. Magnetic field as high as 6T is found
to increase the ferromagnetic character, but the CE-type
antiferromagnetic structure almost remains undisturbed in
$Pr{}_{0.5}Ca_{0.5}Mn_{0.97}Ga_{0.03}O{}_{3}$ compound. A similar
description based on martensitic like scenario on the strain
accomodation in phase separated manganites has been proposed to
explain the step like behavior in M(H) \cite{V. Hardy,V. Hardy-1}.
The appearance of sharp metamagnetic steps in M(H) is attributed
to the presence of different crystallographic unit cell of phase
separated antiferromagnetic and ferromagnetic regions. At low
temperatures, the coexistence of ferromagnetic regions in an
antiferromagnetic matrix results in an elastic constraint effects
at the interface. When the applied magnetic field favoring the
growth of ferromagnetic regions become large enough to overcome
the elastic constraints, it causes a sudden jump in magnetization.
As a consequence of these elastic constraints, the ferromagnetic
regions cannot grow continuously. Fisher et al. have shown that
the quenched disorder induces an inhomogeneous metastable state
and subsequent magnetization jumps in $(SmSr)MnO{}_{3}$ \cite{L.
M. Fisher}. At T = 140K in x = 0.2 sample, the M(H) exhibits a
linear behavior in the paramagnetic regime, while in x = 0.3
sample at 150K a narrow hysteresis loop is still evident,
suggesting the presence of short range ordered ferromagnetic
state. In x = 0.4 sample, the M(H) at 10K shows an extremely
narrow hysteresis loop and a feature of virgin curve lying outside
the envelope curve, while at 140K the M(H) exhibits a linear field
dependence. The behavior of virgin curve lying outside the
envelope curve has also been interpreted in terms of kinetically
arrested first order phase transition and the associated features
like phase coexistence and metastability \cite{A. Banerjee}. The
extent to which the virgin curve lies outside envelope curve is
decided by the initial conditions of two competing phases
(ferromagnetic and antiferromagnetic) in the H-T phase space. A
larger fraction of ferromagnetic phase surviving metastably to
zero field, upon field reversal would cause a larger difference
between virgin and envelope curves \cite{M. A. Manekar-2}. This is
illustrated in figure \ref{Diff-M(H)}, where the difference
between virgin and envelope curves is plotted as a function of Nd
concentration. The difference is maximum in x = 0.3 in comparison
to x = 0.2 and 0.4 samples. This indicates that in x = 0.3 sample
a larger fraction of the ferromagnetic state persists metastably
to zero field upon field reversal. Therefore, at low temperature
all three samples show the presence of ferromagnetic clusters in
an antiferromagnetic matrix, although the fraction / size of these
clusters is significantly larger in x = 0.3 sample.

\begin{figure}[t]
\resizebox{0.50\textwidth}{!}{
\includegraphics{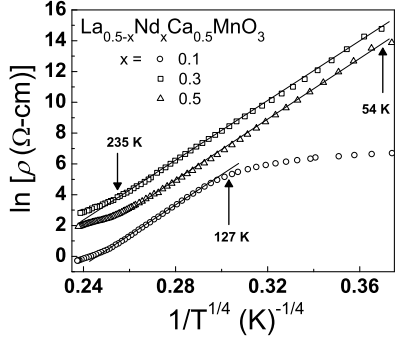}

}\caption{\label{Resistivity}Plot of ln \textgreek{r} versus
temperature $T{}^{-1/4}$ for x = 0.1, 0.3 and 0.5 samples. The
curve for x = 0.5 sample is shifted artificially to enhance
clarity.}

\end{figure}

The temperature dependence of resistivity, $\rho$(T), in x = 0.1,
0.3 and 0.5 compositions are shown in figures \ref{T
dependNd01}(c), \ref{T dependNd03}(c), and \ref{T dependNd05}(c),
respectively. The resistivity data were collected during heating
and cooling cycles. Thermal hysteresis was not observed between
the two cycles. For all the Nd doped samples temperature
dependence of resistivity shows an insulating behavior over the
entire measured temperature range between 50 and 315K. Below 50K
resistivity is too high to be measurable. No significant change in
$\rho$(T) with Nd doping is observed except for x = 0.3 sample. In
this sample reduction in resistivity by nearly two orders of
magnitude is observed in comparison to other Nd doped samples and
is, shown in figure \ref{T dependNd03}(c). The decrease in
$\rho$(T) for x = 0.3 sample can be attributed to the presence of
ferromagnetic regions. In literature various models have been
proposed to explain the conduction mechanism in charge ordered
manganites. In the temperature regime below $\text{T}{}_{CO}$ the
resistivity behavior is best described by Mott\textquoteright{}s
variable range hopping model. We find, the conduction process in
systems with localized effects is described using the variable
range hopping (VRH) model. In this model $\rho$(T) is expressed as
\cite{N. F. Mott}, \[
\rho=\rho_{0}\exp\left(T_{0}/T\right)^{1/4}\] where, $T{}_{0}$ is
the Mott\textquoteright{}s activation energy, and is defined as
$T{}_{0}$ \ensuremath{\approx}
$\frac{21}{K{}_{B}N(E_{F})\xi^{3}}$. In this equation, $k{}_{B}$
is Boltzmann constant, $N\left(E{}_{F}\right)$ is the density of
states at Fermi level, and \textgreek{x} is the localization
length. The characteristic temperature $T{}_{0}$ is related to the
electronic density of states at the Fermi level. The ln $\rho$
versus $T{}^{-1/4}$ plot for samples x = 0.1, 0.3 and 0.5 is shown
in figure \ref{Resistivity}. For clarity the plot for x = 0.5
sample is artificially shifted. The temperature range of fitting
is considerably reduced and shows a deviation from variable range
hopping model below $\sim$130K in x = 0.3 sample. This behavior
may be ascribed to the presence of ferromagnetic interactions in
addition to charge and orbitally ordered antiferromagnetic phase
at low temperatures. No significant change in the value of
$T{}_{0}$ as a function of Nd doping is evident. The value of
$T{}_{0}$ for x = 0.1, 0.3 and 0.5 samples are 0.544 (9)
\texttimes{} $10^{8}K$, 0.87 (2) \texttimes{} $10^{8}K$ and 1.15
(8) \texttimes{} $10^{8}K$, respectively. The value of $T{}_{0}$
for the end compound $Nd{}_{0.5}Ca_{0.5}MnO{}_{3}$ (x = 0.5) is
fairly in agreement with the value
$(T{}_{0}\approx3.4\text{\texttimes}10^{8}K)$ reported in
literature \cite{F. Millange}. In contrast with previously
reported transport and magnetization studies on
$Nd{}_{0.5-x}La_{x}Ca_{0.5}MnO{}_{3}$ series \cite{P. V. Vanitha},
we do not observe metal to insulator transition in these samples
though a reduction in resistivity by two orders of magnitude in x
= 0.3 sample is found.

\subsection{TEM studies}

\begin{figure}[t]
\resizebox{0.50\textwidth}{!}{ \subfloat[at
300K]{\includegraphics[scale=10]{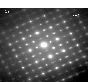}}}
\resizebox{0.50\textwidth}{!}{ \subfloat[at
98K]{\includegraphics[scale=10]{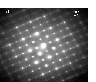}} }
\caption{\label{Diffraction patt Nd03}Selected area diffraction
patterns for $La{}_{0.2}Nd{}_{0.3}Ca_{0.5}MnO{}_{3}$ (x = 0.3)
compound at \textbf{(a)} 300K and \textbf{(b)} 98K. }

\end{figure}

The influence of Nd doping on charge and orbitally ordered
structure in x = 0.3 and 0.4 sample is investigated using TEM
measurements. Electron diffraction patterns and high resolution
lattice images were recorded at various temperatures between 98
and 300K. In both Nd doped samples with x = 0.3 and 0.4, electron
diffraction patterns exhibit identical \textit{Pnma} structure,
showing a typical 90\textdegree{} - domains of orthorhombic
structure in\textit{ Pnma} space group. At 300K diffuse streaks
along a{*} (the reciprocal-space lattice parameter) are observed
and indicate a signature of high temperature charge \textendash{}
orbitally ordered fluctuations, as shown in figure
\ref{Diffraction patt Nd03}(a) for x = 0.3 sample. The presence of
such fluctuations at 300K therefore lead to an approximate
\textit{Pnma} structure. In the temperature region above
$T{}_{CO}$, we have found evidence of short range order using
diffuse neutron scattering studies in related compounds,
$La{}_{0.5}Ca_{0.5-x}Sr{}_{x}MnO{}_{3}$ \cite{I. Dhiman-1}, which
corroborates with the present TEM observation. As temperature is
reduced below 300K down to 220K, no significant difference between
the selected area diffraction patterns is observed. On further
reducing the temperature to 150K, the short range charge ordered
fluctuations is enhanced, leading to an onset of long range
incommensurate charge and orbitally ordered phase. At 98K, the
electron diffraction pattern shows the presence of relatively
intense superlattice spots at positions between the Bragg peaks
corresponding to \textit{Pnma }structure and is shown in figure
\ref{Diffraction patt Nd03}(b). These superlattice spots can be
indexed as q = (1/3 \textendash{} \textgreek{e})a{*}, where
\textgreek{e} is the parameter of incommensurability. By taking
into account the orientation of $e_{g}$ orbitals of $Mn^{3+}$
ions, the lattice parameter along a-axis is doubled leading to
corresponding superlattice spots. Thus it is the orbital ordering,
in combination with charge-ordering, which causes the superlattice
spots \cite{C. H. Chen}. No transition from incommensurate to
commensurate charge ordering structure is observed down to 98K.
Unlike in $La{}_{0.5}Ca_{0.5}MnO{}_{3}$ (x = 0) compound, the
temperature dependent TEM study reveals that the ferromagnetic to
antiferromagnetic transition is accompanied by a incommensurate to
commensurate charge ordering transition at 130K \cite{C. H. Chen}.
In x = 0.4 sample, similar incommensurate charge ordered phase is
observed at 98K. Although, in comparison to x = 0.3 sample the
superlattice spots have much higher intensity, which indicates the
strengthening of charge ordered phase. The occurrence of
incommensuration in Nd doped samples indicate the presence of
charge itinerancy, therefore weakening of antiferromagnetic state.

\subsection{Magnetic Structure}

\begin{figure}[t]
\resizebox{0.50\textwidth}{!}{
\includegraphics{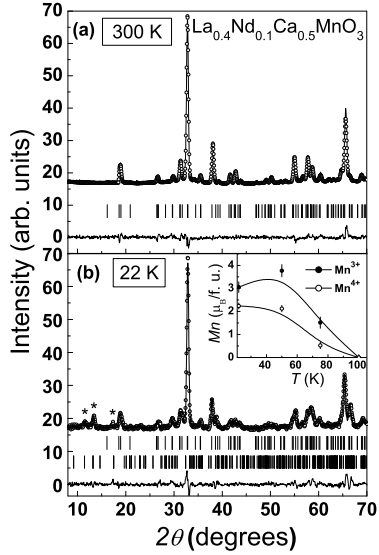}

}\caption{\label{Nd-Nd01}Neutron diffraction pattern recorded on
sample $La{}_{0.4}Nd{}_{0.1}Ca_{0.5}MnO{}_{3}$ (x = 0.1) at
\textbf{(a)} 300 and \textbf{(b)} 22K. This figure is
representative for all the samples having x \ensuremath{\le} 0.5.
Continuous line through the data points is the fitting to chemical
and magnetic structure described in the text. The ({*}) symbols
indicate superlattice reflections having maximum intensity in the
CE-type antiferromagnetic state. The vertical tick marks below the
diffraction pattern in \textbf{(a)} correspond to positions of
Bragg reflections. While in \textbf{(b)} the upper and lower tick
marks indicate the indexing of nuclear and CE-type
antiferromagnetic phases, respectively. The inset of figure
\textbf{(b)} shows the variation of antiferromagnetic site moment
of $Mn{}^{3+}$ and $Mn{}^{4+}$ ions with temperature for sample x
= 0.1. The continuous lines are a guide for the eye.}

\end{figure}

The neutron diffraction pattern of x = 0.1 sample at 300 and 22K
in the angular range of 5\textdegree{} \ensuremath{\le} 2$\theta$
\ensuremath{\le} 70\textdegree{} is shown in figure
\ref{Nd-Nd01}(a) and (b), respectively. This neutron diffraction
pattern is representative for all the samples. On lowering of
temperature below the respective antiferromagnetic ordering
temperatures superlattice reflections are observed, similar to
that of x = 0 sample, indicating the antiferromagnetic nature of
samples \cite{P. G. Radaelli}. The values of refined magnetic
moment on Mn site at 22K and antiferromagnetic ordering
temperature are given in table II. The superlattice reflections
are marked with an asterisks ({*}) symbol in figure
\ref{Nd-Nd01}(b). Particularly, reflections (0, 1, \textonehalf{})
(\textonehalf{}, 1, \textonehalf{}) and (1, 1, \textonehalf{}) are
shown and these characterize the onset of a CE \textendash{} type
antiferromagnetic state. These superlattice reflections are
indexed on a 2a \texttimes{} b \texttimes{} 2c cell in the space
group $P2{}_{1}/m$. In CE \textendash{} type antiferromagnetic
spin structure Mn occupies two distinct sites for $Mn{}^{3+}$ and
$Mn{}^{4+}$ ions. The $Mn{}^{3+}$ and $Mn{}^{4+}$ sublattices are
associated with propagation vector (0, 0, \textonehalf{}) and
(\textonehalf{}, 0, \textonehalf{}), respectively \cite{P. G.
Radaelli}. In this structure, zig zag ferromagnetic chains are
coupled antiferromagnetically within and out of plane. The model
was first proposed by Wollan and Kohler \cite{E.O. Wollan}. Unlike
in the case of Y doped samples \cite{E. Pollert}, the CE - type
antiferromagnetic structure is stable over the entire range of
composition. Though, in the intermediate composition, where the
disorder is maximum, the antiferromagnetic moment is considerably
reduced. The temperature dependence of magnetic moment for x = 0.1
sample is shown in the inset of figure \ref{Nd-Nd01}(b). The
Rietveld refinement of the neutron diffraction pattern at 22K
indicates that the magnetic moment for $Mn{}^{3+}$ and $Mn{}^{4+}$
sites are predominantly oriented along either a or c axis and have
values as 2.5(2)~$\mu{}_{B}$ and 2.2(1)~$\mu{}_{B}$, respectively.
From the present neutron diffraction data no significant change in
$\chi^{2}$ and magnetic R - factor is observed on changing the
orientation of spins between a and c axis, and therefore spins
were constrained to be orientated along a axis. Similarly, the x
and z component of magnetic moment of $Mn{}^{4+}$ ions could not
be refined separately, and therefore the orientation was
constrained to be along c axis. This behavior is similar to the
previously reported neutron diffraction study on
$La{}_{0.5}Ca_{0.5}MnO{}_{3}$ compound by Radaelli et al. \cite{P.
G. Radaelli}. With Nd doping the reorientation of spins from ac
plane to either a or c axis is observed. In addition, no
enhancement of intensity in low angle fundamental Bragg
reflections was visible, which indicates the absence of
ferromagnetic ordering from the present neutron diffraction
experiment. Nevertheless, the M(H) and M(T) study at low
temperature for x = 0.2 and 0.3 samples show signatures for the
presence of ferromagnetism, indicating the moment values are too
small to influence the neutron diffraction data. However, evidence
of ferromagnetic behavior in agreement with M(T) measurements is
found using neutron depolarization measurements, discussed below.
Therefore, M(H) in conjunction with M(T), neutron diffraction, and
neutron depolarization studies indicate the presence of phase
separation behavior, with coexisting ferromagnetic and
antiferromagnetic phases particularly for x = 0.3 composition. For
the end compound $Nd{}_{0.5}Ca_{0.5}MnO{}_{3}$ (x = 0.5) the
refinement of low temperature pattern clearly indicates that the
magnetic moments are oriented along a axis. The value of moments
for $Mn{}^{3+}$ and $Mn{}^{4+}$ are 2.4(1) $\mu{}_{B}$ and 1.8(1)
$\mu{}_{B}$ at 22K, with the transition temperature $T{}_{N}$
\ensuremath{\approx} 125K. In contrast, according to previously
reported neutron diffraction study on
$Nd{}_{0.5}Ca_{0.5}MnO{}_{3}$ compound the $T{}_{N}$
\ensuremath{\approx} 160K, which is higher than our reported value
\cite{F. Millange}.

\subsection{Neutron Depolarization}

The temperature dependence of magnetization in the
anitferromagnetic state in samples x \ensuremath{\le} 0.3 exhibits
an increase in magnetization on lowering of temperature indicating
presence of coexisting ferromagnetic and antiferromagnetic phases.
This behavior is further probed using neutron depolarization
measurements. This technique measures spatial magnetic
inhomogeneities on a length scale from 1000Å to several
micrometers. In the present study, we have measured flipping ratio
R (ratio of the transmitted intensities for two spin states of the
incident neutron spin), which is a measure of the transmitted beam
polarization. R is expressed in the form ,

\[
R=\frac{1-P_{i}DP_{A}}{1+(2f-1)P_{i}DP_{A}}\] where, $P_{i}$ is
the incident beam polarization, $P_{A}$ is the efficiency of the
analyzer crystal, $f$ is the rf flipper efficiency and $D$ is the
depolarization coefficient. In the absence of any depolarization
in sample, $D=1$. $P_{i}D$ is thus the transmitted beam
polarization. As against M(H) measurements, where the sample is
subjected to high field, depolarization measurements provide
information about the presence of ferromagnetic domains in low
magnetic fields and therefore, does not disturb the magnetic
ground state. Hence, depolarization measurement is advantageous to
study the ferromagnetic behavior in samples with coexisting
ferromagnetic and antiferromagnetic phases.

\begin{figure}[t]
\resizebox{0.50\textwidth}{!}{
\includegraphics{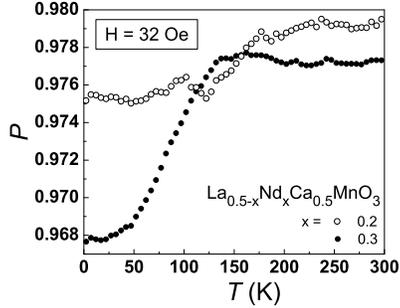}

}\caption{\label{Neutron Polarization}Temperature dependence of
transmitted neutron beam polarization (P) for
$La{}_{0.5-x}$\-$Nd{}_{x}Ca_{0.5}MnO{}_{3}$ (x = 0.2 and 0.3) in H
= 32 Oe is displayed.}

\end{figure}

Figure \ref{Neutron Polarization} shows the temperature dependence
of transmitted neutron beam polarization in a field of 32 Oe, for
x = 0.2 and 0.3 samples. In other samples depolarization is not
observed down to lowest temperature. For x = 0.2 a very small
change in transmitted beam polarization is observed below 150K. In
contrast to x = 0.2 sample, the x = 0.3 compound displays a
significant depolarization below 125K, becoming nearly constant
for T < 50K indicating presence of ferromagnetic domains. The
ferromagnetic nature in x = 0.2 and 0.3 samples observed from
neutron depolarization studies explains the increase in
magnetization in the antiferromagnetic state below 150K observed
in figure \ref{T dependNd05}(c). An estimate of domain size in the
ferromagnetic region is obtained using the expression
$P_{f}=P_{i}exp(-\alpha(d/\delta))<\phi_{\delta}>^{2}$ where,
$P_{f}$ and $P_{i}$ are the transmitted beam and incident beam
polarization, respectively, $\alpha$ is a dimensionless parameter
(= 1/3), $d$ is the sample thickness, $\delta$ is a typical domain
length and the precession angle
$\phi_{\delta}=(4.63\times10^{-10}Oe^{-1}\mathring{A}^{-2})\lambda\delta
B$ \cite{G. Halperin}. The domain magnetization, $B$ is obtained
from the bulk magnetization. This expression is valid in the limit
where domains are randomly oriented and the Larmor phase of
neutron spin due to the internal magnetic field of sample $<2\pi$
over a typical domain length scale. Our measurements were carried
out in low field far away from the saturation field and therefore
satisfy the assumption of this model. The estimated domain size
for x = 0.3 sample at T = 2K is $\sim0.1\mu m$. For higher Nd
doping (x = 0.4), behavior similar to x = 0.2 sample is observed.
The absence or low value of depolarization indicates reduction in
ferromagnetic domain size and / or domain magnetization. This
measurement therefore, gives a clear evidence of ferromagnetic
domains coexisting in the antiferromagnetic state. The absence of
contribution to the Bragg reflections in the low temperature
neutron diffraction studies indicates that the long range
ferromagnetic ordering is absent. Previously, evidence of nano
clusters of ferromagnetic origin in the charge ordered
antiferromagnetic state has been reported \cite{J. C. Loudon}.
Here, we provide the evidence of fairly large ferromagnetic
domains in the charge ordered antiferromagnetic state induced by
disorder. Monte Carlo studies show that nano sized ferromagnetic
clusters in antiferromagnetic phase arise in the electronically
phase separated state where the densities of ferromagnetic and
antiferromagnetic phases are unequal \cite{Dagotto-1}. In the case
where the densities are equal, disorder $\sigma^{2}$ is found to
cause large micrometer size ferromagnetic domains in the
antiferromagnetic matrix, which is evidenced from the present
depolarization studies. Experimentally, coexistence of
ferromagnetic and antiferromagnetic phase has been reported in
$La{}_{0.5}Ca_{0.5}MnO{}_{3}$ \cite{J. C. Loudon,Q. Huang}. As
against, the previous studies, we find that increase in disorder
leads to larger size ferromagnetic domains in antiferromagnetic
matrix, which is in agreement with theoretical studies on phase
separation behavior in charge ordered manganites \cite{Dagotto-1}.

Thus, based on the results presented here it is observed that upon
moderate increase in disorder with Nd doping (x < 0.3), the charge
and orbitally ordered ground state is not fully suppressed.
Simultaneously a ferromagnetic phase is found to grow, leading to
a coexistence of ferromagnetic and antiferromagnetic phases at low
temperature. Moreover, the stepwise behavior evident in M(H) for x
= 0.3 sample corresponds to field induced diminution of charge and
orbitally ordered antiferromagnetic phase and growth of
ferromagnetic phase. These effects are commonly observed in phase
separated systems. Radaelli et al. have shown that for parent
compound (x = 0) the charge and orbitally ordered phase is
accompanied by an increase of ac strain \cite{P. G. Radaelli-1}.
With induced disorder the homogeneous strain field is collapsed
into an inhomogeneous one (as a result of Nd substitution), which
is accompanied by the gradual reduction of charge ordered phase.
Theoretical studies have shown that long range homogenous strain
plays a crucial role in stabilization of charge and orbitally
ordered state in half doped manganites. Ahn et al. have reported
that the presence of uniform strain favors stabilization of the
charge ordered phase \cite{K. H. Ahn}. Recently, the significance
of lattice strain and in turn the influence of disorder on strain
field has been further elucidated in studies of bulk as well as in
thin films of half doped manganites \cite{P. Wagner,W. Prellier,X.
J. Chen,X. J. Chen-1,X. J. Chen-2}. In thin films it has been
shown that epitaxial strain could have a strong effect on charge
ordered state. Therefore, both the interior and exterior strains
are of substantial importance in stabilizing charge and orbitally
ordered phase. For higher Nd doping (x \ensuremath{\ge} 0.4), with
reducing disorder the ferromagnetic phase is fully suppressed
while the charge ordered antiferromagnetic phase is favored. This
behavior is in agreement with increase in Jahn Teller distortions
($\Delta_{JT}$) for x \ensuremath{\ge} 0.4 samples. Hence, a
strong correlation of small disorder with structural and magnetic
behavior is evident.

\section{Conclusion}

We have investigated the influence of small A \textendash{} site disorder
($\sim10{}^{-4}$) on magnetic and transport behavior of charge and
orbitally ordered $La{}_{0.5-x}Nd{}_{x}$\-$Ca_{0.5}MnO{}_{3}$ (0.0
\ensuremath{\le} x \ensuremath{\le} 0.5) series. The disorder as a
function of Nd doping displays a maximum in the range of x = 0.2 -
0.3 compositions. The Nd doped compounds possess orthorhombic structure
in space group \textit{Pnma} at room temperature and on reducing temperature
below $T_{CO}$ exhibit a transition to lower symmetry monoclinic
structure in $P2{}_{1}/m$ space group. The compounds undergo antiferromagnetic
transition between 100 and 125K. Below $T_{N}$ the magnetic structure
is CE-type. Evidence of ferromagnetic domains coexisting with the
CE-type antiferromagnetic ordering is found in the neutron depolarization
and magnetization measurements. The estimated domain size in x = 0.3
sample at T = 2K is $\sim0.1\mu m$. The growth of ferromagnetic phase
exhibits a maximum for x = 0.3 compound. The Jahn \textendash{} Teller
distortion ($\Delta_{JT}$) parameter exhibits a minimum for x = 0.3
compound. The growth of ferromagnetic state results in reduction in
resistivity, though insulator to metal transition is not observed
in the Nd doped samples. Therefore, the evolution magnetic and nuclear
structure at low temperatures as a function of Nd doping exhibit strong
correlation even with small A - site disorder ($\sigma{}^{2}$). \clearpage{}



%

%

\clearpage{}%
\begin{table}
\caption{Structural parameters obtained from Rietveld refinement of neutron
diffraction pattern at 300K for samples $La{}_{0.5-x}Nd{}_{x}Ca_{0.5}MnO{}_{3}$.
The atomic sites are: La/Nd/Ca 4c{[}x, 1/4, z{]}; Mn 4a{[}0, 0, 0{]};
$O{}_{1}$ 4c{[}x, 1/4, z{]}; $O{}_{2}$ 8d{[}x, y, z{]} in Pnma space
group. The symbol $O{}_{1}$ denotes the Oxygen atom along b axis
(apical) and $O{}_{21}$ and $O{}_{22}$ are the two Oxygen atoms
in the ac-plane (equatorial).}

\begin{tabular}{>{\centering}p{2.7cm}>{\centering}p{0.5cm}>{\centering}p{2cm}>{\centering}p{2cm}>{\centering}p{2cm}>{\centering}p{2cm}>{\centering}p{2cm}}
\hline
\noalign{\vskip0.25cm}
Refined parameters &  & x = 0.1 & x = 0.2 & x = 0.3 & x = 0.4 & x = 0.5\tabularnewline
\hline
\noalign{\vskip0.25cm}
a (Å) &  & 5.404 (1)  & 5.399 (1)  & 5.391 (2)  & 5.395 (2)  & 5.3703 (9)\tabularnewline
\noalign{\vskip0.25cm}
b (Å) &  & 7.616 (1)  & 7.607 (1)  & 7.609 (3)  & 7.608 (4)  & 7.564 (1)\tabularnewline
\noalign{\vskip0.25cm}
c (Å) &  & 5.401 (1)  & 5.392 (1)  & 5.393 (2)  & 5.374 (2)  & 5.3595 (9)\tabularnewline
\noalign{\vskip0.25cm}
V ($\mathring{A}\boldsymbol{^{3}}$) &  & 222.30 (8)  & 221.50 (8)  & 221.2 (2)  & 220.6 (1)  & 217.71 (6)\tabularnewline
\noalign{\vskip0.25cm}
<$r_{A}$> &  & 1.1927  & 1.1874  & 1.1821  &  1.1768 & 1.1715 \tabularnewline
\noalign{\vskip0.25cm}
$\sigma{}^{2}\times10{}^{-4}$ &  & 3.8 & 3.92 & 3.41 & 2.35 & 0.723\tabularnewline
\noalign{\vskip0.25cm}
La/Nd/Ca & x  & 0.020 (1)  & 0.018 (1)  & 0.014 (2)  & 0.026 (2)  & 0.0292 (8)\tabularnewline
 & z  & 0.490 (2)  & 0.500 (5)  & 0.484 (3)  & 0.496 (4)  & 0.493 (2)\tabularnewline
\noalign{\vskip0.25cm}
$O{}_{1}$  & x  & 0.493 (2)  & 0.495 (3)  & 0.499 (4)  & 0.481 (2)  & 0.488 (1)\tabularnewline
 & z  & 0.559 (2)  & 0.565 (2)  & 0.557 (3)  & 0.569 (3)  & 0.564 (1)\tabularnewline
\noalign{\vskip0.25cm}
$O{}_{2}$ & x  & 0.278 (1)  & 0.277 (1)  & 0.287 (2)  & 0.281 (2)  & 0.284 (1)\tabularnewline
 & y  & 0.0319 (8)  & 0.030 (9)  & 0.032 (2)  & 0.031 (1)  & 0.0349 (7)\tabularnewline
 & z  & 0.220 (2)  & 0.222 (2)  & 0.213 (3)  & 0.219 (2)  & 0.2145 (9)\tabularnewline
\noalign{\vskip0.25cm}
$Mn-O{}_{1}$ (Å) &  & 1.931 (1)  & 1.935 (2)  & 1.927 (3)  & 1.941(3)  & 1.923(1)\tabularnewline
\noalign{\vskip0.25cm}
$Mn-O{}_{21}$ (Å)  &  & 1.930 (7)  & 1.931 (8)  & 1.943 (13)  & 1.933 (10)  & 1.930 (5)\tabularnewline
\noalign{\vskip0.25cm}
$Mn-O{}_{22}$ (Å)  &  & 1.946 (7)  & 1.935 (8)  & 1.943 (14)  & 1.932 (11)  & 1.937 (5)\tabularnewline
\noalign{\vskip0.25cm}
$Mn-O{}_{1}-Mn$ (\textdegree{}) &  & 160.91 (7)  & 158.74 (8)  & 161.1 (1)  & 157.1 (1)  & 159.20 (6)\tabularnewline
\noalign{\vskip0.25cm}
$Mn-O{}_{2}-Mn$ (\textdegree{}) &  & 160.5 (3)  & 161.6 (4)  & 157.7 (6)  & 160.1 (4)  & 157.7 (2)\tabularnewline
\hline
\end{tabular}
\end{table}
\clearpage{}%
\begin{table}
\caption{Structural parameters obtained from Rietveld refinement of neutron
diffraction pattern at 22K for samples $La{}_{0.5-x}$\-$Nd{}_{x}Ca_{0.5}MnO{}_{3}$.
The atomic sites are: La/Nd/Ca 4c{[}x, 1/4, z{]}; Mn 4a{[}0, 0, 0{]};
$O{}_{1}$ 4c{[}x, 1/4, z{]}; $O{}_{2}$ 8d{[}x, y, z{]} in Pnma space
group. The symbol $O{}_{1}$ denotes the Oxygen atom along b axis
(apical) and $O{}_{21}$ and $O{}_{22}$ are the two Oxygen atoms
in the ac-plane (equatorial).}
\begin{tabular}{>{\centering}p{2.7cm}>{\centering}p{0.5cm}>{\centering}p{2cm}>{\centering}p{2cm}>{\centering}p{2cm}>{\centering}p{2cm}>{\centering}p{2cm}}
\hline
\noalign{\vskip0.25cm}
Refined parameters &  & x = 0.1 & x = 0.2 & x = 0.3 & \textbf{x} = 0.4 & x = 0.5\tabularnewline
\hline
\noalign{\vskip0.25cm}
a (Å) &  & 5.415 (1) & 5.396 (1) & 5.399 (2) & 5.406 (1) & 5.386 (1)\tabularnewline
\noalign{\vskip0.25cm}
b (Å) &  & 7.5203 (9) & 7.556 (1) & 7.554 (2) & 7.519 (1) & 7.475 (4)\tabularnewline
\noalign{\vskip0.25cm}
c (Å) &  & 5.435 (1) & 5.421 (1) & 5.403 (2) & 5.401 (1) & 5.379 (1)\tabularnewline
\noalign{\vskip0.25cm}
V ($\mathring{A}\boldsymbol{^{3}}$) &  & 221.35 (7) & 221.04 (8) & 220.4 (1) & 219.56 (9) & 216.57 (7)\tabularnewline
\noalign{\vskip0.25cm}
La/Nd/Ca & x  & 0.020 (1) & 0.024 (2) & 0.016 (2) & 0.036 (1) & 0.0292 (9)\tabularnewline
 & z  & 0.496 (3) & 0.492 (2) & 0.485 (2) & 0.489 (1) & 0.491 (1)\tabularnewline
\noalign{\vskip0.25cm}
$O{}_{1}$ & x & 0.488 (2) & 0.487 (2) & 0.503 (3) & 0.483 (1) & 0.487 (1)\tabularnewline
 & z & 0.561 (1) & 0.562 (1) & 0.557 (2) & 0.559 (1) & 0.562 (1)\tabularnewline
\noalign{\vskip0.25cm}
$O{}_{2}$ & x & 0.273 (1)  & 0.271 (2) & 0.285 (3) & 0.279 (1) & 0.281 (1)\tabularnewline
 & y & 0.0357(4) & 0.0329 (7) & 0.0316 (9) & 0.0348 (6) & 0.0354 (5)\tabularnewline
 & z & 0.223 (3) & 0.223 (2) & 0.211 (3) & 0.224 (1) & 0.216 (1)\tabularnewline
\noalign{\vskip0.25cm}
$\Delta_{JT}\times10^{-4}$ &  & 1.6 (4)  & 1.2 (6) & 0.8 (3) & 1.0 (4) & 1.2 (4)\tabularnewline
\noalign{\vskip0.25cm}
$T{}_{N}$ (K) &  & 100 & 100 & 100 & 125 & 125\tabularnewline
\noalign{\vskip0.25cm}
$M$ ($\mu_{B}$) &  &  &  &  &  & \tabularnewline
\noalign{\vskip0.25cm}
$\mu_{x}$ ($Mn^{3+}$) &  & 2.5 (2)  & 0.8 (4)  & 1.7 (3)  & 2.6 (2)  & 2.5 (2)\tabularnewline
\noalign{\vskip0.25cm}
$\mu_{x}$ ($Mn^{4+}$) &  & 2.2 (1)  & 0.9 (1)  & 0.6 (2)  & 2.2 (1)  & 1.8 (1)\tabularnewline
\noalign{\vskip0.25cm}
$Mn-O{}_{1}$ (Å) &  & 1.909 (1) & 1.921 (2) & 1.914 (2) & 1.908 (1) & 1.899 (1)\tabularnewline
\noalign{\vskip0.25cm}
$Mn-O{}_{21}$ (Å)  &  & 1.926 (8) & 1.913 (10) & 1.933 (16) & 1.954 (7) & 1.926 (7)\tabularnewline
\noalign{\vskip0.25cm}
$Mn-O{}_{22}$ (Å)  &  & 1.968 (6) & 1.961 (10) & 1.956 (2) & 1.927 (8) & 1.950 (7)\tabularnewline
\noalign{\vskip0.25cm}
$Mn-O{}_{1}-Mn$ (\textdegree{}) &  & 159.97 (5) & 159.18 (7) & 161.32 (7) & 160.14 (5) & 159.38 (5)\tabularnewline
\noalign{\vskip0.25cm}
$Mn-O{}_{2}-Mn$ (\textdegree{}) &  & 160.2 (3) & 161.6 (4) & 158.3 (7) & 159.8 (3) & 158.2 (3)\tabularnewline
\hline
\end{tabular}
\end{table}
\clearpage{}


\begin{thebibliography}{61}
\bibitem{C N R Rao}C. N. R. Rao and B. Raveau, Colossal Magnetoresistance,
Charge Ordering, and Related Properties of Manganese Oxides World
Scientific, 1998, Singapore.

\bibitem{Dagotto}E. Dagotto, Nanoscale Phase Separation and Colossal
Magnetoresistance, Springer Series in Solid State Physics Vol. 136
Springer, 2003, Berlin.

\bibitem{J. B. Goodenough}J. B. Goodenough, in Handbook on the Physics
and Chemistry of Rare Earth, edited by K. A. Gschneidner, Jr., J.-C.
Bunzli, and V. K. Pecharsky Elsevier Science, Vol. 33, 2003, Amsterdam.

\bibitem{Y. Tokura}Y. Tokura, Rep. Prog. Phys. 69 (2006) 797.

\bibitem{L. M. Rodriguez-Martinez}L. M. Rodriguez-Martinez and J.
P. Attfield, Phys. Rev. B 54 (1996) R15622, L. M. Rodriguez-Martinez
and J. P. Attfield, Phys. Rev. B 58 (1998) 2426.

\bibitem{H. Kuwahara}H. Kuwahara, Y. Moritomo, Y. Tomioka, A. Asamitsu,
M. Kasai, R. Kumai and Y. Tokura, Phys. Rev. B 56 (1997) 9386.

\bibitem{J. Wolfman}J. Wolfman, Ch. Simon, M. Hervieu, A. Maignan,
B. Raveau, J. Solid State Chem. 123 (1996) 413.

\bibitem{L. S. Ling}L. S. Ling, S. Tan, L. Pi and Y. H. Zhang, Europhys.
Lett. 79 (2007) 47008.

\bibitem{Md. Motin Seikh}Md. Motin Seikh, L. Sudheendra, and C. N.
R. Rao, J. Solid State Chem. 177 (2004) 3633.

\bibitem{J. Curiale}J. Curiale, C. A. Ramos, P. Levy, R. D. Sanchez,
F. Rivadulla, and J. Rivas, Physica B 354 (2004) 47.

\bibitem{P. V. Vanitha}P. V. Vanitha and C. N. R. Rao, J. Phys.:
Condens. Matter 13 (2001) 11707.

\bibitem{D. Zhu}D. Zhu, X. Tan, P. Cao, F. Jia, X. Ma, and Y. Lu,
J. Appl. Phys. 105 (2009) 063914.

\bibitem{S. Yang}S. Yang, J. Zhong, J. Miao, J. Yuan, B. Xu, L. Cao,
X. Qiu, B. Zhao, Z. Xie, and L. Zhao, Physica B 370 (2005) 99.

\bibitem{A. Moreo}A. Moreo, S. Yunoki and E. Dagotto, Science 283
(1999) 2034.

\bibitem{Dagotto-1}E. Dagotto, T. Hotta, A. Moreo, Phys. Rep. 344
(2001) 1.

\bibitem{J. van den Brink}J. van den Brink, G. Khaliullin and D.
Khomskii, Phys. Rev. Lett. 83 (1999) 5118.

\bibitem{L. Brey}L. Brey, Phys. Rev. B 71 (2005) 174426.

\bibitem{A. Moreo-1}A. Moreo, M. Mayr, A. Feiguin, S. Yunoki and
E. Dagotto, Phys. Rev. Lett. 84 (2000) 5568.

\bibitem{K. Pradhan}K. Pradhan, A. Mukherjee and P. Majumdar, Phys.
Rev. Lett. 99 (2007) 147206.

\bibitem{A. Das}A. Das, P. D. Babu, Sandip Chatterjee and A. K. Nigam,
Phys. Rev. B 70 (2004) 224404,

\bibitem{P. D. Babu}P. D. Babu, A. Das, S. K. Paranjpe, Solid State
Commun. 118 (2001) 91.

\bibitem{I. Dhiman}I. Dhiman, A. Das, P. K. Mishra, and L. Panicker,
Phys. Rev. B 77 (2008) 094440,

\bibitem{I. Dhiman-1}I. Dhiman, A. Das, R. Mittal, Y. Su, A. Kumar,
and A. Radulescu, Phys. Rev. B 81 (2010) 104423.

\bibitem{I. Dhiman-2}I. Dhiman, A. Das and A. K. Nigam, J. Phys.:
Condens. Matter 21 (2009) 386002.

\bibitem{C. Autret-Lambert}C. Autret-Lambert, Z. Jirak, M. Gervais,
N. Poirot, F. Gervais, N. Raimboux, P. Simon, F. Boure, and G. Andr,
Chem. Mater. 19 (2007) 5222.

\bibitem{S. Savitha Pillai}S. Savitha Pillai, G. Rangarajan, N. P.
Raju, A. J. Epstein and P. N. Santhosh, J. Phys.: Condens. Matter
19 (2007) 496221.

\bibitem{P. G. Radaelli}P. G. Radaelli, D. E. Cox, M. Marezio and
S. -W. Cheong, Phys. Rev. B 55 (1997) 3015.

\bibitem{F. Millange}F. Millange, S. de Brion, and G. Chouteau, Phys.
Rev. B 62 (2000) 5619.

\bibitem{J. Rodriguez-Carvajal}J. Rodriguez-Carvajal, Physica B 192
(1993) 55.

\bibitem{B. B. van Aken}B. B. Van Aken, O. D. Jurchescu, A. Meetsma,
Y. Tomioka, T. Tokura, and T. T. M. Palstra, Phys. Rev. Lett. 90 (2003)
066403.

\bibitem{G Maris}G. Maris, V. Volotchaev and T. T. M. Palstra, New
J. Phys. 6 (2004) 153.

\bibitem{R. D. Shannon}A-site cationic radii for ninefold coordination
in oxides were taken from R. D. Shannon, Acta Crystallogr., Sect.
A: Cryst. Phys., Diffr., Theor. Gen. Crystallogr., A32 (1976) 751.

\bibitem{J. B. Goodenough-1}J. B. Goodenough, Phys. Rev. 100 (1955)
564.

\bibitem{A. Daoud-Aladine}A. Daoud-Aladine, J. Rodríguez-Carvajal,
L. Pinsard-Gaudart, M. T. Fernández-Díaz, and A. Revcolevschi, Phys.
Rev. Lett. 89 (2002) 097205.

\bibitem{L. Wu}L. Wu, R. F. Klie, Y. Zhu, and Ch. Jooss, Phys. Rev.
B 76 (2007) 174210.

\bibitem{C. H. Chen}C. H. Chen and S. -W. Cheong, Phys. Rev. Lett.
76 (1996) 4042.

\bibitem{J. Rodr=0000EDguez-Carvajal}J. Rodríguez-Carvajal, M. Hennion,
F. Moussa, A. H. Moudden, L. Pinsard, and A. Revcolevschi, Phys. Rev.
B 57 (1998) 3189(R).

\bibitem{E. Fertman}E. Fertman, A. Beznosov, D. Sheptyakov, V. Desnenko,
M. Kajnakova, A. Feher and D. Khalyavin, J Magn. Magn. Mater. 321
(2008) 316.

\bibitem{T. Vogt}T. Vogt, A. K. Cheetham, R. Mahendiran, A. K. Raychaudhuri,
R. Mahesh and C. N. R. Rao, Phys. Rev. B 54 (1996) 15303.

\bibitem{P. Murugavel}P. Murugavel, C. Narayana, A. K. Sood, S. Parashar,
A. R. Raju and C. N. R. Rao, Europhys. Lett. 52 (2000) 461.

\bibitem{J. C. Loudon}J. C. Loudon, N. D. Mathur, and P. A. Midgley,
Nature London 420 (2002) 19.

\bibitem{L. Ghivelder }L. Ghivelder and F. Parisi, Phys. Rev. B 71
(2005) 184425.

\bibitem{C. R. Serrao}C. R. Serrao, A. Sundaresan, and C. N. R. Rao,
J. Phys.: Condens. Matter 19 (2007) 496217.

\bibitem{C. Yaicle}C. Yaicle, C. Martin, Z. Jirak, F. Fauth, G. Andre,
E. Suard, A. Maignan, V. Hardy, R. Retoux, M. Hervieu, S. Hebert,
B. Raveau, Ch. Simon, D. Saurel, A. Brulet, and F. Bouree, Phys. Rev.
B 68 (2003) 224412.

\bibitem{V. Hardy}V. Hardy, S. Hebert, A. Maignan, C. Martin, M.
Hervieu, and B. Raveau, J. Magn. and Magn. Mater. 264 (2003) 183.

\bibitem{V. Hardy-1}V. Hardy, C. Yaicle, S. Hebert, A. Maignan, C.
Martin, M. Hervieu, and B. Raveau, J. Appl. Phys. 94 (2003) 5316.

\bibitem{L. M. Fisher}L. M. Fisher, A. V. Kalinov, I. F. Voloshin,
N. A. Babushkina, D. I. Khomskii, Y. Zhang, T. T. M. Palstra, Phys.
Rev. B 70 (2004) 212411.

\bibitem{A. Banerjee}A. Banerjee, K. Mukherjee, K. Kumar and P. Chaddah,
Phys. Rev. B 74 (2006) 224445, A. Banerjee, A. K. Pramanik, K. Kumar
and P. Chaddah, J. Phys.: Condens. Matter 18 (2006) L605.

\bibitem{M. A. Manekar-2}M. A. Manekar and S. B. Roy, Eur. Phys.
J. B 64 (2008) 19.

\bibitem{N. F. Mott}N. F. Mott and E. A. Davis, Electronic Processes
in Non Crystalline Materials $2{}^{nd}$ edn. 1979 (Oxford: Clarendon).

\bibitem{E.O. Wollan}E.O. Wollan and W.C. Kohler, Phys. Rev. 100
(1955) 545.

\bibitem{E. Pollert}E. Pollert, S. Krupicka, E. Kuzmicova, J. Phys.
Chem. Solids 43 (1982) 1137.

\bibitem{G. Halperin}G. Halperin and T. Holstein, Phys. Rev. 59 (1941)
960 , R. W. Erwin, J. Appl. Phys. 67 (1990) 5229.

\bibitem{Q. Huang}Q. Huang, J. W. Lynn, R. W. Erwin, A. Santoro,
D. C. Dender, V. N. Smolyaninova, K. Ghosh, and R. L. Greene, Phys.
Rev. B 61 (2000) 8895.

\bibitem{P. G. Radaelli-1}P. G. Radaelli, D. E. Cox, M. Marezio,
S.-W. Cheong, P. E. Schiffer, and A. P. Ramirez, Phys. Rev. Lett.
75 (1995) 4488.

\bibitem{K. H. Ahn}K. H. Ahn, T. Lookman, and A. R. Bishop, Nature
(London) 428 (2004) 401.

\bibitem{P. Wagner}P. Wagner, I. Gordon, A. Vantomme, D. Dierickx,
M. J. Van Bael, V. V. Moshchalkov, and Y. Bruynseraede, Europhys.
Lett. 41 (1998) 49.

\bibitem{W. Prellier}W. Prellier, A. Biswas, M. Rajeswari, T. Venkatesan,
and R. L. Greene, Appl. Phys. Lett. 75 (1999) 397.

\bibitem{X. J. Chen}X. J. Chen, S. Soltan, H. Zhang, and H.-U. Habermeier,
Phys. Rev. B 65 (2002) 174402.

\bibitem{X. J. Chen-1}X. J. Chen, H.-U. Habermeier, and C. C. Almasan,
Phys. Rev. B 68 (2003) 132407.

\bibitem{X. J. Chen-2}X. J. Chen, H. -U. Habermeier, H. Zhang, G.
Gu, M. Varela, J. Santamaria, and C. C. Almasan, Phys. Rev. B 72 (2005)
104403.
\end{thebibliography}
\end{document}